\documentclass[preprint2]{aastex62}
\usepackage{amsmath}
\usepackage{bm}
\graphicspath{{./}{figures/}}

\received{--}
\revised{--}
\accepted{--}
\shorttitle{Wave generation and heat flux suppression}
\shortauthors{Roberg-Clark et al.}

\begin{document}

\title{Wave generation and heat flux suppression in astrophysical plasma systems}

\author[0000-0001-5280-2644]{G.~T.~Roberg-Clark}
\affil{Department of Physics, University of Maryland, College Park, MD 20740, USA}

\author[0000-0002-9150-1841]{J.~F.~Drake}
\affil{Department of Physics, University of Maryland, College Park, MD 20740, USA}
\affil{Institute for Research in Electronics and Applied Physics, University of Maryland, College Park, MD 20742, USA}
\affil{Institute for Physical Science and Technology, University of Maryland, College Park, MD 20742, USA}
\affil{Joint Space-Science Institute (JSI), College Park, MD 20742, USA}
\email{drake@umd.edu}

\author[0000-0002-5435-3544]{M.~Swisdak}
\affil{Department of Physics, University of Maryland, College Park, MD 20740, USA}
\affil{Institute for Research in Electronics and Applied Physics, University of Maryland, College Park, MD 20742, USA}
\affil{Joint Space-Science Institute (JSI), College Park, MD 20742, USA}
\email{Swisdak@umd.edu}

\author[0000-0002-1510-4860]{C.~S.~Reynolds}
\affil{Institute of Astronomy, Madingley Road, Cambridge, CB3 OHA, UK}
\email{csr12@ast.cam.ac.uk}

\begin{abstract}

Heat flux suppression in collisionless plasmas for a large range of plasma $\beta$ is explored using two-dimensional particle-in-cell simulations with a strong, sustained thermal gradient. We find that a transition takes place between whistler-dominated (high-$\beta$) and double-layer-dominated (low-$\beta$) heat flux suppression. Whistlers saturate at small amplitude in the low beta limit and are unable to effectively suppress the heat flux. Electrostatic double layers suppress the heat flux to a mostly constant factor of the free streaming value once this transition happens. The double layer physics is an example of ion-electron coupling and occurs on a scale of roughly the electron Debye length. The scaling of ion heating associated with the various heat flux driven instabilities is explored over the full range of $\beta$ explored. The range of plasma-$\beta$s studied in this work makes it relevant to the dynamics of a large variety of astrophysical plasmas, including the intracluster medium of galaxy clusters, hot accretion flows, stellar and accretion disk coronae, and the solar wind.

\end{abstract}



\section{Introduction} \label{sec:intro}

The microphysics of thermal conduction in weakly collisional, weakly magnetized astrophysical plasmas is an important topic that remains to be fully elucidated. It has long been recognized that the Spitzer-H\"arm thermal conductivity \citep{Spitzer1953,Spitzer1962}, while appropriate for describing thermal conduction in highly collisional fluids, may not be the correct prescription for weakly collisional or collisionless plasmas in which the mean free path is larger than or comparable to the scale size of the system. This has motivated study of the microturbulence produced by kinetic plasma instabilities and how it can modify thermal transport in weakly collisional astrophysical systems.

Two environments in which collisionless thermal conduction may play a significant role are the solar wind (for which $\beta = 8\pi n T / B^{2} \sim 1$) and the intracluster medium (ICM) of galaxy clusters (where $\beta \sim 100$). Instabilities driven by heat-flux-carrying particle distributions in the solar wind and their impact on thermal conduction have been studied since at least the 1970's and remain under active investigation (e.g. \cite{Hollweg1974,Gary1975,Hollweg1976,Gary1978,Gary1994,Gary2000,Saeed2017a,Saeed2017b,Shaaban2018a,Horaites2018a}). An advantage of studying thermal conduction in the solar wind is the availability of in-situ measurements at $1$ AU, at which the measured value of $\beta$ can in fact vary significantly, reaching even $\beta=100$ \citep{Bale2013}. Measured heat fluxes at $1$ AU can therefore serve as a proxy for transport in distant, high-$\beta$, weakly collisional plasmas such as the ICM. 

Thermal conduction can be crucial to the dynamics of the ICM, a fact that has manifested itself in models of, for example, thermodynamic stability in galaxy clusters (e.g. \cite{Zakamska2003a,Kim2003a,Ruszkowski2010,Fang2018a}), AGN feedback (e.g. \cite{Kannan2017a,Yang2016b,Tang2018a,Reynolds2015,Bambic2018a}), and the propagation of sound waves in cluster cores (e.g. \cite{Fabian2005a,Zweibel2018a}). It also is a necessary ingredient of fluid-type instabilities such as the heat-flux-driven buoyancy instability (HBI) \citep{Quataert2008a} and magnetothermal instability (MTI) \citep{Balbus2000a}. These applications provide a strong impetus to understand the detailed microphysics of thermal conduction in weakly collisional plasmas and how it might affect the large-scale ICM.
 
\subsection{Whistler Heat Flux Suppression} \label{sec:Whistler}

Previous work \citep{Roberg-Clark2016,Roberg-Clark2018,Komarov2018a} has explored the suppression of electron thermal conduction by electromagnetic whistler waves. Whistlers are observed in a large variety of space plasma environments, including the solar wind \citep{Beinroth1981,Lacombe2014}, the Earth's bowshock \citep{Wilson2016,Oka2017}, and the Earth's radiation belts \citep{Mozer2014,Agapitov2015}. They are also often self-consistently generated in local kinetic simulations of systems such as astrophysical shocks \citep{Riquelme2011,Guo2017} and hot accretion flows \citep{Riquelme2016,Riquelme2017,Sironi2015a,Sironi2015b}. It was shown in \cite{Roberg-Clark2016} that large-amplitude whistlers driven unstable by an electron heat flux can strongly inhibit thermal conduction. The mechanism is scattering via the overlap of cyclotron resonances \citep{Smith1975,Karimabadi1992a} and requires that the whistlers propagate obliquely to the local magnetic field \citep{Pistinner1998}.  

Results from 2D particle-in-cell (PIC) simulations with an imposed electron thermal gradient (\cite{Roberg-Clark2018,Komarov2018a}) suggest that the maximum electron thermal conduction parallel to the local magnetic field scales roughly like $1/\beta$ for $\beta > 1$. Such a scaling is consistent with earlier theoretical models \citep{Gary1994,Levinson1992,Gary2000,Pistinner1998} and some observations of electron heat flux in the solar wind \citep{Gary1998,Gary1999}. The $1/\beta$ limit has recently been confirmed by observations of solar wind heat flux measured by the WIND Spacecraft at $1$AU \citep{Tong2018} for $\beta \sim 1 - 6$. However, it is clear from the data shown in \cite{Tong2018} that a transition to a different limiting heat flux is at play when $\beta \lesssim 2$. In the following section we describe a means by which heat flux can be limited at low $\beta$.

\subsection{Double Layer Heat Flux Suppression} \label{sec:DL}

Recent PIC simulation results presented in a series of papers by Li, Drake \& Swisdak (hereafter LDS) \citep{Li2012,Li2013,Li2014} identified a transport suppression mechanism mediated by electrostatic double layers (DLs) in the $\beta \sim 1$ regime. LDS modeled the outward propagation of a localized source of hot electrons produced at a coronal looptop during a solar flare. In their scheme an initial gradient in the electron temperature drove a parallel electron heat flux. The resulting hot electron current drove a return current carried by the cold background electrons that penetrated into the hot source region. The return current then coupled to the ions via the Buneman instability \citep{Buneman1958,Volokitin1982}, which in its nonlinear stage of evolution produced a localized electrostatic potential identified as a DL \citep{Block1978,Singh1987,Raadu1988}. Finite mass (i.e. non-stationary) ions were required in the LDS simulations to capture the interaction between the return current electrons and the ambient ions that led to the formation of the DL. 

A DL functions effectively as a mobile parallel-plate capacitor in a plasma, maintaining a potential drop across two layers of opposite charge whose separation is of the order of the Debye length $\lambda_D$ \citep{Block1978}. In the LDS simulations the self-sustaining and long-lived DL (and in some cases multiple DLs, \citep{Li2014}) propagated in the direction of the return current. Heat flux was inhibited as the DLs reflected hot electrons propagating from the source region and continued to accelerate a cold return current opposing the motion of hot electrons. However, suppression of heat flux was modest since only particles with energy less than the DL electrostatic potential were reflected and confined within the source region, while higher-energy hot electrons were allowed to pass through the potential with minimal reduction in energy \citep{Li2012}.

LDS's results suggest that DL physics may play a role in magnetized plasmas with sustained temperature gradients such as those simulated in \cite{Roberg-Clark2018} and \cite{Komarov2018a} but in a more modest $\beta \sim 1$ regime relevant to systems such as the solar wind and coronal looptops, where it has been speculated that turbulent magnetic fluctuations may be responsible for confining hot electrons during flares \citep{Kontar2014}. Other systems of interest might include low-luminosity accretion flows and their coronae, since results from GRMHD simulations (e.g. \cite{Sadowski2013a}) imply that $\beta$ may be of order unity at high latitudes near the coronae of black holes \citep{Sironi2015b}.
We find that DLs indeed play a central role in the $\beta \sim 1$ regime and in the following sections present a series of PIC simulations that reveal the transition from whistler heat flux suppression to suppression by double layers.
 
\section{Numerical Method and Simulation Parameters} \label{sec:numerical}

We carry out two-dimensional (2D) simulations using the PIC code $\tt{p3d}$ \citep{Zeiler2002} to model thermal conduction along an imposed temperature gradient in a
magnetized, collisionless plasma with open boundaries as in \citep{Roberg-Clark2018}. $\tt{p3d}$
calculates particle trajectories using the relativistic Newton-Lorentz
equations and the electromagnetic fields are advanced using Maxwell's
equations. The ends of the simulation domain act as thermal reservoirs
at two different temperatures $T_{h} > T_{c}$ separated by a distance
$L_{x}$, forming a temperature gradient $T' \equiv
(T_{h}-T_{c})/L_{x}$ and driving a heat flux. An initially uniform
magnetic field $\mathbf{B_{0}}=B_{0} \mathbf{\hat{x}}$ threads the
plasma along the gradient and is free to evolve in time. The initial
particle distribution function is chosen to model the free-streaming
of particles from each thermal reservoir and has the form \small
\begin{multline}\label{eqn:1}
f(\mathbf{v},t=0) = f_{h} + f_{c} \\ = \frac{n_{0}}{\pi^{3/2}} \left( \frac{ e^{-v^2/v_{Th}^2}}{v_{Th}^3}\theta(v_{\parallel}) + \frac{e^{-[(v_{\parallel} + v_{d})^2 + v_{\perp}^2]/v_{Tc}^2}}{v_{Tc}^3(1+\text{erf}(v_{d}/v_{Tc}))}\theta(-v_{\parallel}) \right)
\end{multline}
\normalsize where $n_0$ is the initial density, $\theta$ is the
Heaviside step function, $v_{T}=\sqrt{2T/m}$ is the thermal speed, and the parallel and perpendicular directions are with respect to $\mathbf{B_0}$. The cold particles are given a parallel drift speed $v_{d}$ to ensure zero net current ($\langle v_{\parallel} \rangle =0$) in the initial state while the error function $\text{erf}(v_{d}/v_{Tc})$ makes the density of hot and cold particles equal.

In the simulations presented here ion positions and velocities are evolved in time, allowing double layers to form. Ions, with mass ratio $m_{i}/m_{e}=1600$, are initialized with a Maxwellian distribution of temperature $T_{i0}=T_{eh}/2$ and are re-injected into the domain using the above scheme with equal temperatures $T_{ih}=T_{ic}=T_{i0}$. 

The simulations scan a  range of $\beta_{e0h}=8\pi n_{0}T_{eh}/B^{2}_{0}$ from $32$ to $1/4$ in factors of $2$ ($32$,$16$,$8$,...,$1/4$) using an electron temperature ratio $T_{eh}/T_{ec} = 10$. The simulation domain lengths are $L_{x}=L_{0}=164 \: d_{e}$ and $L_{y}=L_{0}/2$, where $d_{e} = c/\omega_{pe}$ is the electron skin depth and $\omega_{pe}=(4\pi n_{0}e^{2}/m_{e})^{1/2}$ is the electron plasma frequency. The characteristic velocity of whistlers depends on the wavelength but has an upper limit that scales with the electron Alfv\'en speed  $V_{A,e}=d_{e}\Omega_{e0}$. However, we normalize electron heat fluxes to the free-streaming value $q_{0} = n_{0}v_{T_{eh}}T_{eh}$ as in \cite{Roberg-Clark2018}.

Since DLs are generated near the cold reservoir and propagate towards the hot thermal reservoir at $x=0$, we stop the simulation before the DL conduction front reaches the hot boundary and significantly impacts plasma injection. For $\beta_{e0h}=1$ this corresponds to a time of $t \Omega_{e0} \sim 7200$. Other parameters in the simulations are $\omega_{pe}/\Omega_{e0} = 5 \sqrt{\beta_{e0h}}$, and $T_{eh}/(m_{e}c^{2})=0.02$, which sets $v_{Th}/c = 1/5$ such that electrons are mostly non-relativistic. Each simulation uses $560$ particles per species per cell and has a grid of $4096$ by $2048$ cells.

\section{Simulation Results} \label{sec:results}

\subsection{Whistler regime} \label{sec:whistler}

\begin{figure*}[ht!]
\plotone{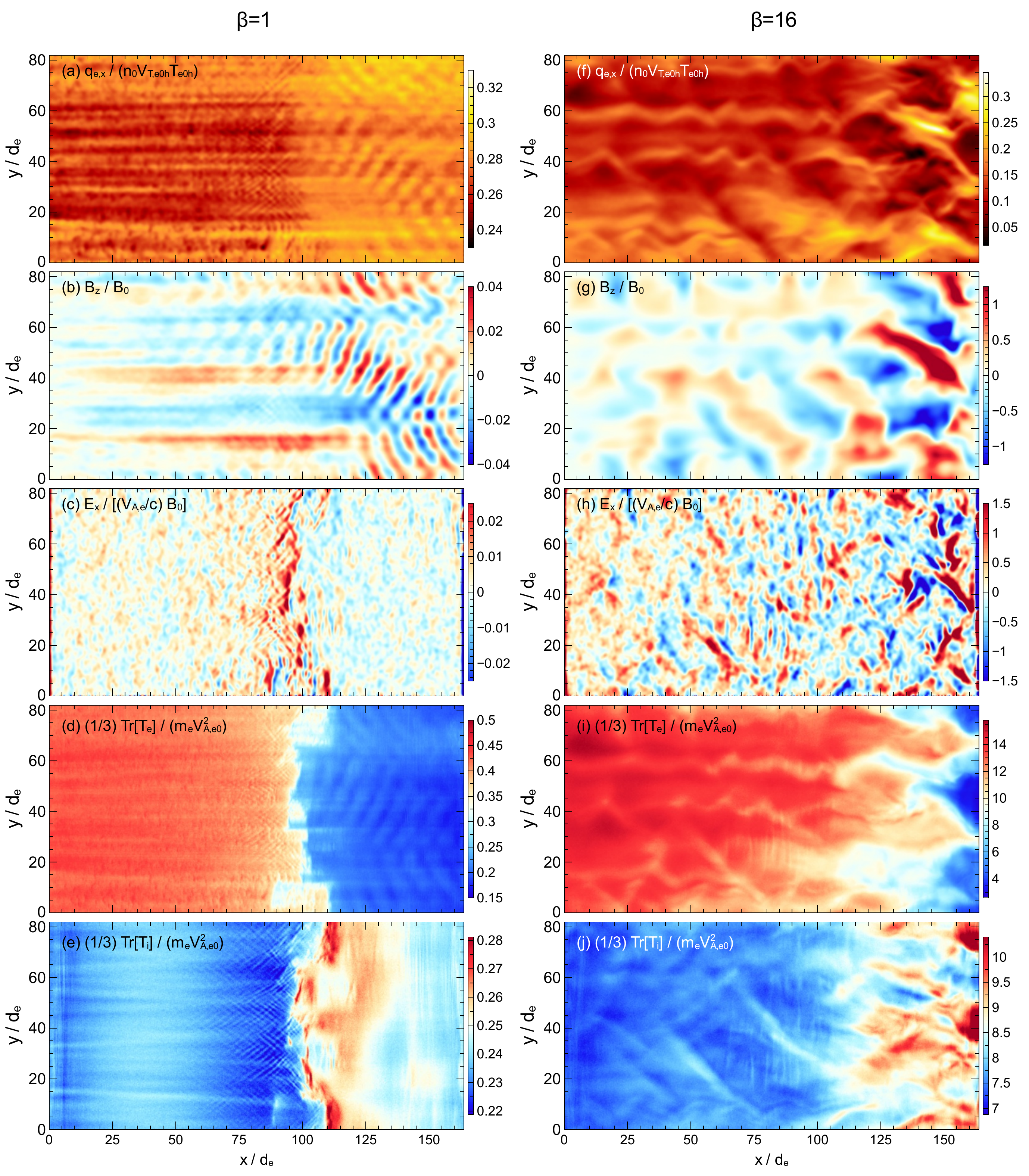}
\caption{2D plots from the $\beta_{e0h} =1$ [(a)-(e)] and $16$ [(f)-(j)] simulations at $t\Omega_{e0}=3640$ and $910$, respectively. (a) Suppression of $q_{ex}$ by DLs located between $x/d_{e} \sim 85 - 115$ that propagate towards the hot reservoir at $x=0$. (b) Out-of-plane $B_{z}$ showing elongated magnetic structures mostly to the left of the DLs and oblique whistlers mostly to the right of the DLs. (c) Parallel electric field $E_{x}$ showing a DL conduction front (bright red spots). (d) One third of the trace of the electron temperature tensor $T_{e}$. (e) Same as (d) but for ions showing substantial heating to the right of the DLs. (f) Strong suppression of $q_{ex}$ by whistlers. (g) Large-amplitude whistler fluctuations in $B_{z}$. (h) Parallel whistler electric fields in $E_{x}$ (near the cold reservoir at $x = L_{x}$) and turbulence associated with a small DL conduction front near $x = L_{x}/2$. (i) 2D structure of $T_{e}$ in whistler fields. (j) Heating of ions by whistlers.}
\label{fig:1}
\end{figure*}

Figure \ref{fig:1} contrasts the dynamics of the $\beta_{e0h}=1$ [(a)-(e)] and $16$ [(f)-(j)] simulations. In the case of $\beta_{e0h}=16$ the results are similar to the large $\beta_{e}$ ($32,64,$ and $128$) runs of \cite{Roberg-Clark2018}. Strong heat flux suppression below the free-streaming value takes place throughout the simulation domain [Fig. \ref{fig:1}(f)] and is associated with scattering of electrons by large-amplitude oblique whistler fluctuations $\delta{B}/B_{0}\sim 1$ in the out-of-plane $B_{z}$ in \ref{fig:1}(g).

Figure \ref{fig:1}(j) shows substantial heating of ions, correlated with the turbulent whistler fields near the cold reservoir at $x = L_{x}$. In some locations the peak ion temperature is $40 \%$ larger than $T_{i0}$. The ions are scattered by the spatially localized, $\rho_{e0h}(=\sqrt{\beta_{e0h}}d_{e})$ scale whistler electric fields and have an unmagnetized response to the turbulence $(\rho_{i0} \gg \rho_{e0h})$. The electric field fluctuations $E_x$ associated with these whistlers are shown in Fig.~\ref{fig:1}(h). To estimate the rate of ion heating we calculate an energy diffusion coefficient for homogeneous whistler turbulence 
\begin{equation}\label{eqn:3}
D_{W} = \frac{\langle \Delta v_{i}^{2} \rangle}{t} = \frac{q^{2}_{i}}{m^{2}_{i}} \int_{0}^{\infty} d\tau \langle \delta E_{x}(0) \delta E_{x}(\tau) \rangle.
\end{equation} 
Similar expressions for electron diffusion in whistler turbulence are given in \cite{Keenan2016a}. The dynamics of (\ref{eqn:3}) can also be framed as a resonant diffusive process which is essentially Landau damping [see e.g. \cite{Sturrock1966b} and \cite{Morales1974}]. Assuming isotropic turbulence and dropping factors of order unity we get $D_{W} \sim (q_{i}^2/m^{2}_{i})\tau_{c} \langle \delta E^{2} \rangle$, where $\tau_{c}$ is the correlation time of the electric field, which we take to be $ \tau_{c} = \rho_{e0h}/v_{p} \sim \beta_{e0h}/\Omega_{e0}$.

The electric field fluctuations $E_{x}$ [Fig. \ref{fig:1}(h)] are of the order of $(V_{A,e}/c) \: B_{0}$, a factor of $\sqrt{\beta_{e0h}}=4$ larger than the electric field of a parallel propagating whistler with phase speed $v_{p}=v_{Teh} / \beta_{e0h}$,
\begin{equation}\label{eqn:2}
\delta E = \frac{v_{p}}{c} \delta B \sim \frac{v_{Teh}}{c\beta_{e0h}}B_{0}.
\end{equation}
However, we still measure phase speeds of the order $v_{Teh} / \beta_{e0h}$ in the $\beta_{e0h}=16$ simulation (not shown). The crux is that these whistlers are oblique, with $\mathbf{k}$ nearly parallel to $\delta \mathbf{E}$. To be consistent with Faraday's Law, $\mathbf{k} \times \delta \mathbf{E} = (\omega/c) \delta \mathbf{B}$, the electric field is enhanced by a factor of $1/\sin(\theta)$ relative to (\ref{eqn:2}), where $\theta$ is the angle between $\mathbf{k}$ and $\delta \mathbf{E}$. We nonetheless take expression (\ref{eqn:2}) to be correct and use it as a reasonable lower bound for $\delta E$ in the large-amplitude and high-$\beta$ regime.

Using (\ref{eqn:2}) leads to $D_{W} \sim B_{0}^{2}/(4\pi n_{0}m_{i}) \: \Omega_{i0}$. Multiplying by $(1/2)n_{0}m_{i}$ we arrive at an ion heating rate
\begin{equation}\label{eqn:4}
\frac{\partial{W_{i}}}{\partial t} = \frac{1}{2}n_{0}m_{i} D_{W} \sim \frac{B_{0}^{2}}{8\pi} \Omega_{i0}.
\end{equation}
Inserting numbers from the $\beta_{e0h}=16$ simulation (\ref{eqn:4}) yields a net heating of roughly $10\%$ of $T_{i0}$ by $t\Omega_{e0}=910$. Spatially averaging $(1/3)\text{Tr}[T_{i}]$ over all $y$ and the region $x = 0.7 \: L_{x}$ to $x = 0.9 \: L_{x}$ (where large-amplitude whistlers are most prevalent) yields ion heating of around $5\%$ of $T_{i0}$, which is within a factor of $2$ of the analytic estimate. Equation (\ref{eqn:4}) suggests local ion heating can occur on the very rapid time scale of ion cyclotron motion in the high-$\beta$ ICM when strong whistler turbulence is present. Over long time scales in the ICM the damping of the whistlers associated with ion heating can be offset by a net positive growth rate from the heat flux instability. In the regime of large collisionless heat flux the damping from ions does not significantly impact whistler stability (see Fig. \ref{fig:4} and section \ref{sec:saturation} as well as \cite{Komarov2018a}). The regime of smaller heat fluxes and marginal heat flux instability will be investigated in future work.

\subsection{DL regime}\label{sec:DL}

In the $\beta_{e0h} = 1$ simulation a thermal conduction front, containing several DLs (as in \cite{Li2014}) in fig. \ref{fig:1}(c), divides the simulation domain into "hot" ($x/d_{e} \lesssim 100$) and "cold" ($x/d_{e} \gtrsim 100$) regions. The conduction front originates near the cold thermal reservoir and propagates into the hot region (evidence for this is shown in figure 2). The DLs are the dark-red, mostly vertical structures in fig. \ref{fig:1}(c) around $x/d_{e}=100$ and have parallel wavelengths of roughly $10 \: \lambda_{Deh}$ \citep{Li2012}, where $\lambda_{Deh}=V_{Teh}/(\sqrt{2}\omega_{pe})$ is the hot electron Debye length. These wavelengths are consistent with the unstable modes of the Buneman instability \citep{Buneman1958}. Two DLs are present at the bottom of figure \ref{fig:1}(c) around $y/d_{e}=5$ and $x/d_{e} \sim 87$ and $x/d_{e} \sim 110$ \citep{Li2014}. The potential jump across the DL front is $e\Phi_{DL} \sim 0.4 \: T_{e0h}$, consistent with the results and analytic predictions of \citep{Li2013}.

The DLs are dynamical structures that tend to break up and reform over time, developing nontrivial structure in the perpendicular ($y$) direction. The perpendicular length scale for the DLs is roughly $30 \: \lambda_{Deh}$, consistent with results from previous 2D PIC simulations of double layers (e.g. \cite{Barnes1985,Main2013}), although the mechanism that sets the angle between the DL electric field and $B_{0}$ (and hence $k_{\perp}/k_{\parallel}$) remains an open question \citep{Ergun2004}, which we do not address in this paper.

Heat flux suppression for $\beta_{e0h}=1$ [fig. \ref{fig:1}(a)] is moderate, with a minimum heat flux of roughly $0.24 \: q_{0}$ in the hot region. Mostly rightward-propagating oblique whistlers ($kd_{e}\sim 1$) with small saturation amplitude ($\delta B/B_{0} \sim 0.04$) are present in the cold region, while the hot region contains no whistlers and instead develops elongated magnetic structures $B_{z}$ with $k_{\perp}d_{e} \sim 1$ and $k_{\parallel}d_{e} \sim 0.07$ [fig. \ref{fig:1}(b)]. Note the reversals in the sign of $B_{z}$ along the $y$ direction. These structures seem to merge with the whistler fluctuations at the approximate location of the conduction front at $x/d_{e} \sim 100$ and are discussed in more detail in section \ref{sec:kaw}. 

The elongated magnetic structures do not appear to significantly impact the heat flux [fig. \ref{fig:1}(a)], which is nonetheless modulated by thin streams with even smaller perpendicular wavelength in the hot region. Rather, it is the DLs, which reflect hot electrons and accelerate the return current, that reduce the overall heat flux in the hot region. Evidence for reflected hot particles and the return current are shown in figure 3. The small-amplitude whistlers in the cold region have a negligible impact on the heat flux (\ref{fig:1}a) unlike in the high $\beta_{e}$ regime.

Figure \ref{fig:1}(d) shows the trace of the electron temperature tensor in $(x,y)$, indicating a sharp discontinuity in $T_{e}$ around the conduction front at $x/d_{e} \sim 100$ with temperatures somewhat less than $T_{e0h}$ at the hot reservoir and $T_{e0h}/3$ at the cold reservoir ($T_{e0h}=0.5 \: m_{e}V_{A,e0}^{2}$). The ion temperature [fig. \ref{fig:1}(e)] reveals local heating to be roughly $12 \%$ above the injection temperature $T_{i0}$ (note that $T_{i0}= 0.25 \: m_e V_{A,e0}^{2}$). 

A simple estimate for ion heating by the DL front can be obtained by calculating the change in kinetic energy of ions accelerated by the DL. We assume ions are initially cold and that they gain a velocity $v_{\Phi} = \sqrt{2e\Phi_{DL}/m_{i}}$ as they cross the DL in the $+\hat{x}$ direction (the velocity of the DL is small compared with $v_{\Phi}$). The result is a distribution with two beam-like populations (one with $v_{x}=0$ and the other with $v_{x}=v_{\Phi}$) in the cold region. The effective thermal energy of the ions is $(3/2)\Delta T_{i}=(1/2)m_{i}v_{\Phi}^{2}/4$. With $e\Phi_{DL} \sim (2/5)T_{eh}$, we find
\begin{equation}\label{eqn:5}
\Delta T_{i} \sim \frac{1}{15}T_{eh},
\end{equation}
which predicts a heating of about $12\%$ above $T_{i0}$, close to the observed heating in fig. \ref{fig:1}(e). 

\subsection{Generation and propagation of the DLs}

\begin{figure}
\plotone{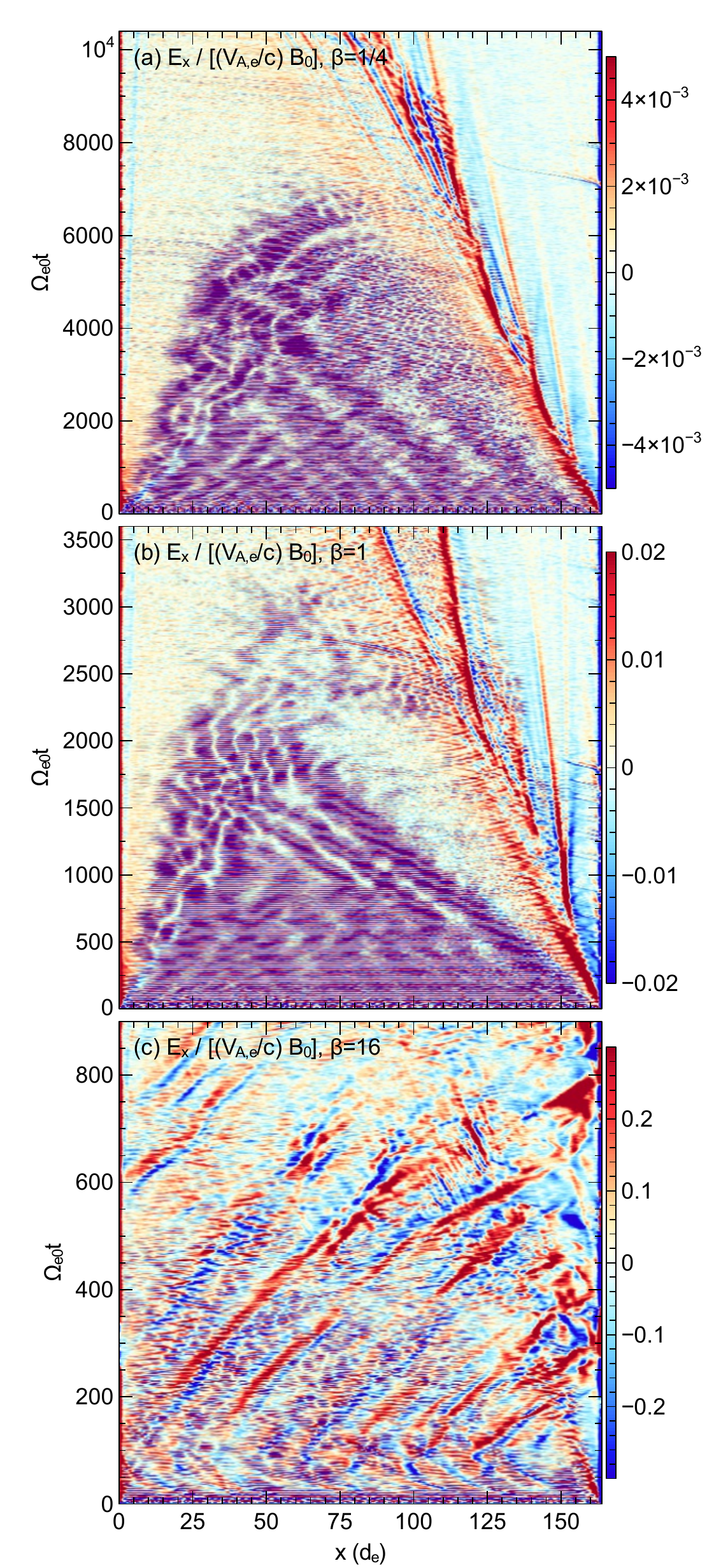}
\caption{Spacetime diagrams of $E_{x}$ at $y=3.2 \: d_{e}$ for three values of $\beta_{e0h}$ showing DL formation at the cold reservoir $x=L_{0}$. (a) $\beta_{e0h}=1/4$: several strong DLs form and travel towards the hot reservoir, emitting slow-moving ion-acoustic shocks to the right. (b) $\beta_{e0h}=1$: Similar to (a) but with two primary DLs forming at later stages of the simulation. (c) $\beta_{e0h}=16$: $E_{x}$ is dominated by rightward-traveling whistler fluctuations that overtake the weak DL.}
\label{fig:2}
\end{figure}

Figure \ref{fig:2} shows a spacetime diagram of the electric field $E_{x}$, at a cut along $y=L_{y}/2$ for the $\beta_{e}=1/4$ [\ref{fig:2}(a)], $1$ [\ref{fig:2}(b)], and $16$ [\ref{fig:2}(c)] simulations. At early times the initial electron distribution function $f_0$ (\ref{eqn:1}) relaxes, generating short-wavelength ($k\lambda_{Deh}\sim 1$), large-amplitude electron acoustic waves resulting from two-stream instability between the hot and cold electron populations \citep{Gary1987}, visible for all $x$ near $t=0$. These waves damp out early on in the simulation and do not impact the subsequent dynamics. Also present at early times are plasma waves (purple lines merging with the electron acoustic modes) which propagate from the cold region to the hot region and have $kd_{e} \sim 3$. While these waves are fully resolved in time in the simulation, data outputs are less frequent and show a strongly alternating pattern. These waves tend to damp out by the middle of each simulation while confined to the region to the left of the DL.

In \ref{fig:2}(a-c) a DL forms at early times near the cold reservoir at $x=L_{x}$ and then propagates toward the hot reservoir at roughly the acoustic speed $V_{s} \sim \sqrt{(T_{e0h}+T_{i0})/m_{i}}$ \citep{Li2013}.  A single DL can be distinguished in the electric field $E_{x}$ by a large positive amplitude on its leading edge to the left, followed by a small negative leg on the right \citep{Li2012,Li2013,Li2014}. Each DL grows in amplitude as it is continually fed by return currents but stabilizes itself by emitting an ion acoustic shock towards the cold reservoir, which can be identified by the nearly vertical straight lines with positive amplitude (red) to the right of the DLs in fig. \ref{fig:2} \citep{Li2013,Li2014}. The shocks quench DL growth by reflecting cold return current electrons before they reach the DLs \citep{Li2013,Li2014}. However, the shocks damp over time [see equation (\ref{eqn:5})], leading to subsequent DL growth and shock re-emission. 

Increasing $\beta_{e0h}$ reduces the number and robustness of the DLs [several are visible in fig. \ref{fig:2}(a), two in \ref{fig:2}(b) and one only at early time in \ref{fig:2}(c)]. Increasing $\beta_{e0h}$ also increases the strength of whistlers. For $\beta_{e0h}=16$ [\ref{fig:2}(c)] the DL front is essentially overwhelmed by whistler electric fields (seen as rightward propagating lines in the whole domain for $t\Omega_{e0} \gtrsim 100)$) as the waves propagate toward the hot reservoir.  The weakness of the DLs in the high $\beta_{e}$ regime is a result of strong scattering of return current electrons by whistlers (not shown), which removes the drive mechanism for Buneman instability and hence suppresses DL formation and propagation.

\subsection{Electron dynamics in the hot and cold regions}

\begin{figure*}[ht!]
\plotone{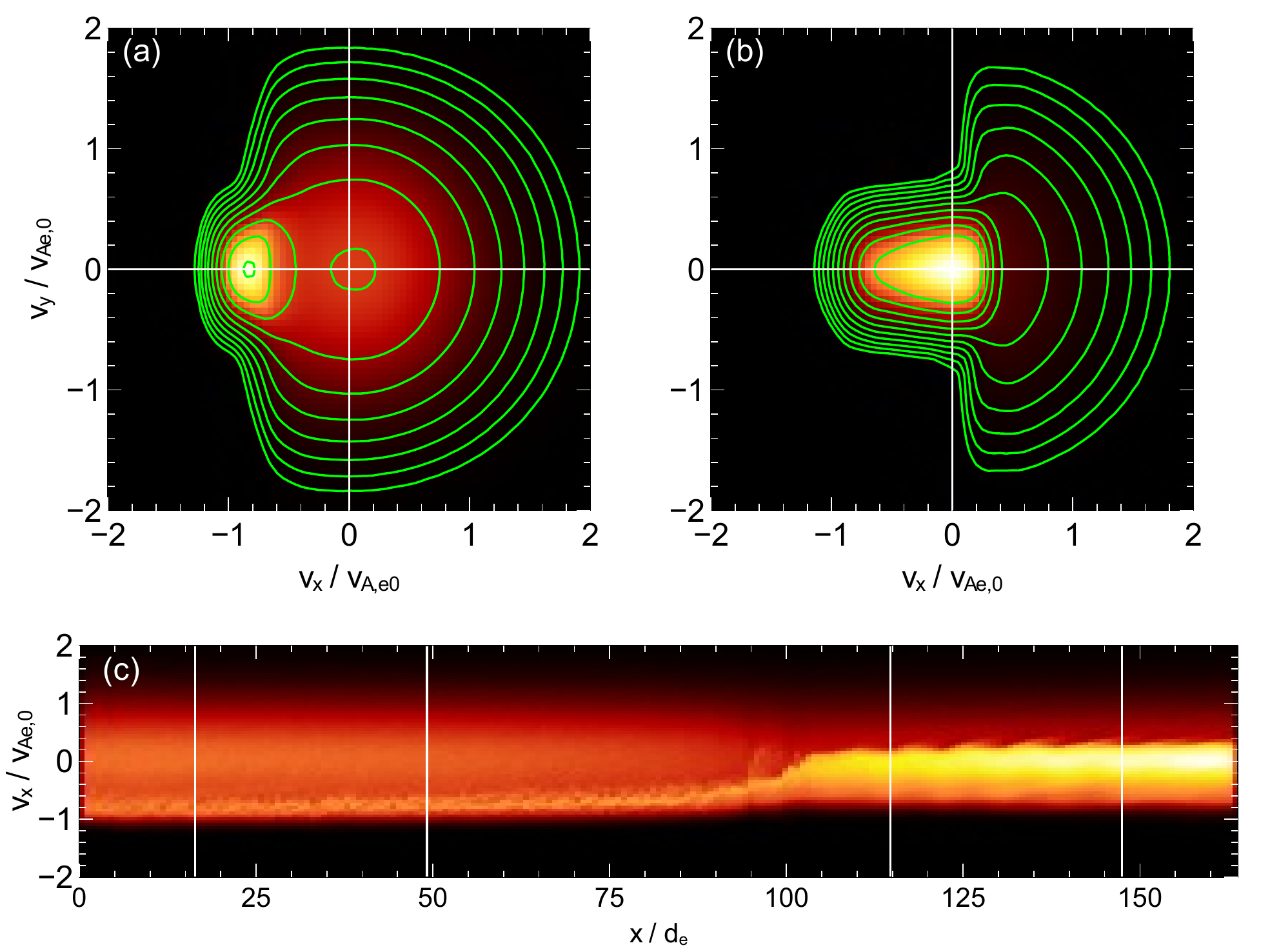}
\caption{Plots from the $\beta_{e0h}=1$ simulation taken at $t\Omega^{-1}_{e0}=3640$. (a) Electron distribution function $f_{e}$ (background color) and ln$(f_{e})$ (green contours) in $v_{x}-v_{y}$ space averaged over the spatial region $x=0.1 \: L_{x}$ to $x=0.3 \: L_{x}$ (the "hot" region) and all $y$. (b) Same as (a) but averaged over $x=0.7$ to $0.9 \: L_{x}$ (``cold" region). (c) $f_{e}$ in the phase space $x-v_{x}$ averaged over a thin strip in $y$ near $y=L_{y}/2$. The sections in $x$ over which $f_{e}$ was averaged to produce (a) and (b) are indicated by vertical white stripes.}
\label{fig:3}
\end{figure*}

The electron distribution function for lower $\beta_{e0h}$ is similar to $f_{0}$ in Eq.~(\ref{eqn:1}) but has been altered by the DL potential. Figures \ref{fig:3}(a) and (b) show color plots of $f_{e}(v_{x},v_{y})$ in the $\beta_{e0h}=1$ simulation at $t\Omega_{e0}=3640$, averaged over spatial regions to the left (hot region) and right (cold region) of the DL indicated by vertical white lines in fig. \ref{fig:3}(c). The contours of ln($f_{e}$) are overlaid in green. In fig. \ref{fig:3}(a) an accelerated return current beam (the bright spot to the left of $v_{x}=0$) is present while the contours indicate that the distribution function is smooth and roughly circular for $v_{x}>0$. The discontinuity (i.e. hot/cold interface) in the distribution of injected electrons $f_{0}$ at $v_{x}=0$ has been filled in by hot particles with $-\sqrt{2 e\Phi_{\Phi}/m_{e}} \lesssim v_{x} \lesssim 0$ reflected by the DL potential. In figure \ref{fig:3}(b) the hot/cold interface is shifted slightly to the right but remains near $v_{x}=0$. Figure \ref{fig:3}(c) shows the phase space profile $f_{e}(x,v_{x})$ of electrons averaged over a thin vertical extent in $y$ near $y=L_{y}/2$. The presence of return currents is evident for all $x$, which focus into a narrow beam for $x/d_{e} \lesssim 100$. The narrowing of the overall width in $v_{x}$ from hot to cold regions reveals the gradient in the electron temperature shown in figure \ref{fig:1}(d).

\subsection{Electron thermal conduction}

\begin{figure*}
\plotone{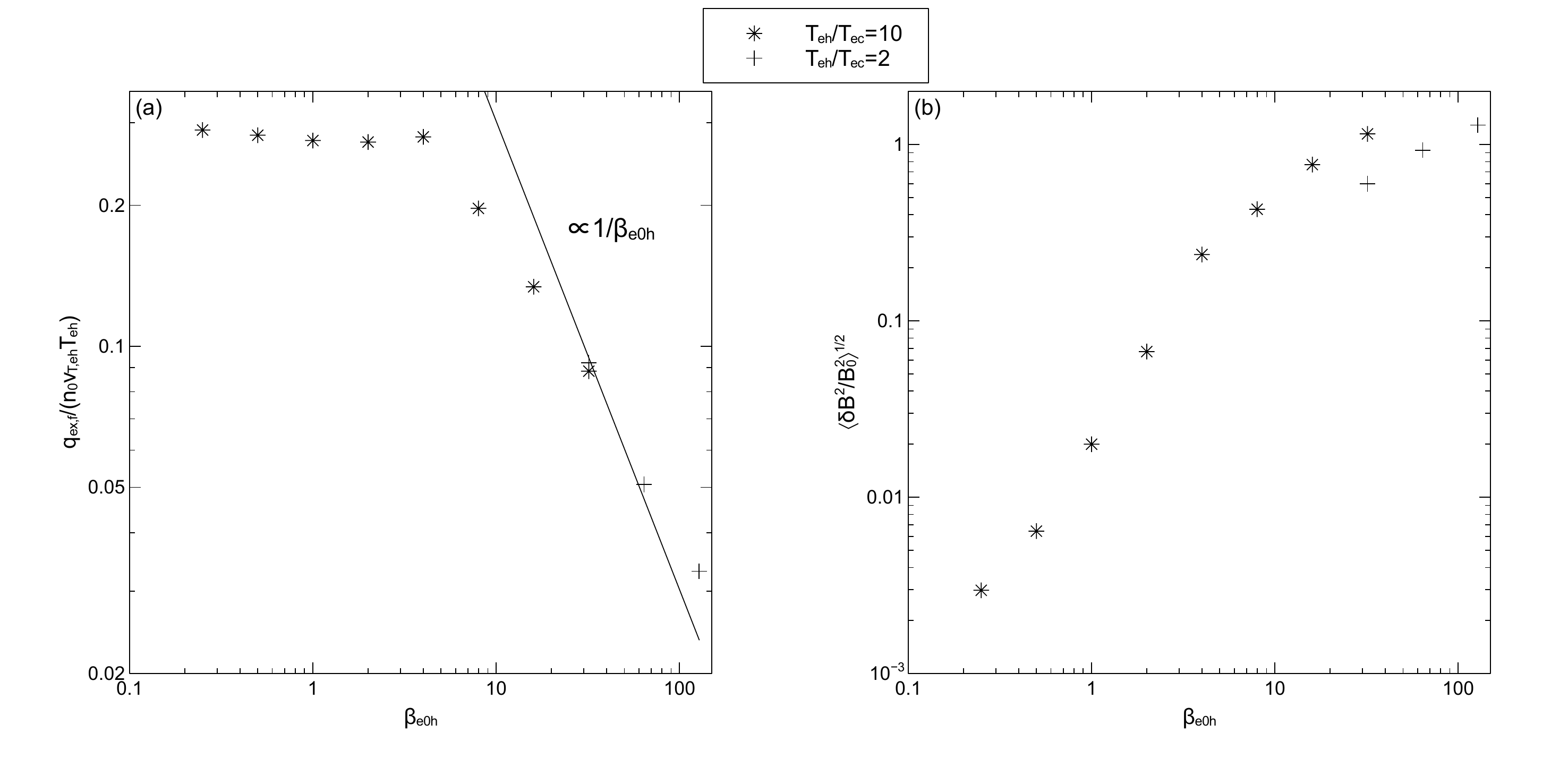}
\caption{(a) Plot of the final measured heat flux from simulations as a function of $\beta_{e0h}$, normalized to the free-streaming value $q_{e0}=n_{0}v_{Teh}T_{eh}$. Data from \cite{Roberg-Clark2018} is shown as plus signs while new simulation data is shown in asterisks. A line proportional to $1/\beta_{e0h}$ (as in \cite{Roberg-Clark2018}) is overlaid. (b) Late-time perturbed magnetic field amplitudes $\langle\delta{B^{2}}/B^{2}_{0}\rangle^{1/2}$ as a function of $\beta_{e0h}$, obtained by averaging over $x=0.7 \: L_{x}$ to $x=0.9 \: L_{x}$ and all of $y$. The extent in $x$ was chosen to always be to the right of the DL conduction front for the new simulations with mobile ions.}
\label{fig:4}
\end{figure*}

Figure \ref{fig:4}(a) shows the late time electron heat flux as a function of $\beta_{e0h}$, including the new simulations with $T_{eh}/T_{ec}=10$ and mobile ions as well as three simulations taken from \cite{Roberg-Clark2018} (with $T_{eh}/T_{ec}=2$ and stationary ions). The $1/\beta_{e0h}$ scaling (shown as a solid line overlaid) matches the data for $\beta_{e0h} \gtrsim 16$. Since the data at $\beta_{e0h}=32$ with and without mobile essentially overlaps, we confirm that the ions have little impact on the heat flux at high $\beta_{e0h}$. A rollover to a roughly constant value $q/q_{0} \sim .29$ in the final heat flux occurs for $\beta_{e0h} \leq 4$ and is set by the DL potential as shown by the following calculation. Assuming $e\Phi_{DL} / T_{eh} = 0.4$, that return current electrons are a $T=0$ beam with velocity $v_{b}/V_{Teh} = -0.8$ [from fig. \ref{fig:3}(a)], that hot electrons with $|v_{\parallel}| \leq \sqrt{2e\Phi/m_{e}}$ contribute no current or heat flux owing to reflection by the DL, and imposing zero net current we calculate the heat flux in the hot region to be $q_{ex}/n_{0}V_{Teh}T_{eh} \sim 0.33$, which is close to the measured value of $0.29$.

\subsection{Saturated whistler amplitudes}\label{sec:saturation}

We document the strength of the whistler heat flux instability over the range of our  high $\beta_{e0h}$ simulations, by showing in figure \ref{fig:4}(b) the spatial average of the perturbed magnetic field, $\langle \delta B^{2} / B^{2}_{0} \rangle^{1/2}$. The data from each simulation is from late time such that the turbulence has saturated. In simulations with DLs, the DL front has propagated into the center of the simulation domain. The spatial region for averaging is $x=0.7 \: L_{x}$ to $x=0.9 \: L_{x}$ and all of $y$. For the high $\beta_{e0h}$ runs this is where the whistler turbulence is strongest and for the low $\beta_{e0h}$ runs with DLs it is the ``cold" region. The general trend in figure \ref{fig:4}(b) is increasing magnetic fluctuation amplitude as $\beta_{e0h}$ is increased. Two distinct regimes are evident in this figure \ref{fig:4}(b): the region of strong heat flux instability ($\beta_{e0h} \gtrsim 4$), where the characteristic saturated whistler amplitude approaches $B_{0}$ as $\beta_{e0h}$ is increased; and the region of weak (or nonexistent) heat flux instability for $\beta_{e0h} \lesssim 1$, where the magnetic fluctuations mostly consist of the elongated structures mentioned in section \ref{sec:DL}. We now review the basic physics of heat flux instability and describe these two regimes.

\subsubsection{Strong heat flux instability}

In the high-$\beta_{e0h}$ regime the oblique whistler goes to long wavelengths, $k \rho_{e} \sim 1$, and the Landau resonance at $v_{x}=v_{p} \sim v_{Te}/\beta_{e0h} \ll v_{Teh}$ aligns with the hot/cold interface in $f_{e}$ (\ref{eqn:1}). Once amplitudes (and hence nonlinear trapping widths) are large enough, the Landau and cyclotron resonances can simultaneously overlap \citep{Roberg-Clark2016}, allowing whistler turbulence to isotropize the hot electron distribution function about the characteristic whistler phase speed. As the hot electrons are scattered about $v_{p}$ they continue to release free energy and drive whistler growth. To estimate the whistler saturation amplitude in this regime, we calculate the free energy difference $\Delta W = W_{h,\text{final}} - W_{h,\text{initial}}$, between the final, isotropized distribution and an initial half-Maxwellian distribution of hot electrons moving with $v_{x} > 0$. The details of the calculation are shown in Appendix A. 

To maintain zero net current in the final state, cold return current electrons must also be displaced in the $v_{x}-v_{y}$ plane to cancel the current $\sim v_{p}$ of the isotropized hot particles. In these simulations the return current drift is reduced by re-injection at the cold thermal reservoir as the current of hot particles is suppressed by whistler scattering. We therefore neglect the energetics of cold return current electrons in this calculation, noting that in general an induced electric field could also maintain current neutrality \citep{Komarov2018a,Ramani1978,Levinson1992}.

We find that, to lowest order in $v_{p}/v_{Teh}$, the energy lost by the hot electron half-Maxwellian as it scattered is proportional to $v_{p}/v_{Teh}$ (\ref{eqn:A6}). For a linear whistler wave at high $\beta$, the energy content is mostly in magnetic field fluctuations and particle kinetic energy and electric fields can be neglected. This should hold true for large-amplitude whistlers as well. Taking $\Delta W \sim -n_{0}T_{e0h}v_p/v_{e0h}$ with $v_p = v_{Teh} / \beta_{e0h}$ and $\beta_{e0h}=8\pi n_{0}T_{eh}/(B^{2}_{0})$, we obtain
\begin{equation}\label{eqn:8}
W_{\text{whistler}} \sim \frac{\langle \delta B^{2} \rangle}{8\pi} \sim \Delta W_{h} \sim \frac{B^{2}_{0}}{8\pi},
\end{equation}
i.e., $\delta B \sim B_{0}$. This explains why there seems to be an order unity ($B_{0}$) limit on the size of magnetic fluctuations in the simulations in fig. \ref{fig:4}(b). The limit comes from free energy constraints on the hot electron distribution function as it is scattered by the large-amplitude whistlers. Coincidentally the $\delta B \sim B_{0}$ at high $\beta$ result was also predicted in \cite{Komarov2018a} through a resonance broadening estimate.

\begin{figure*}
\plotone{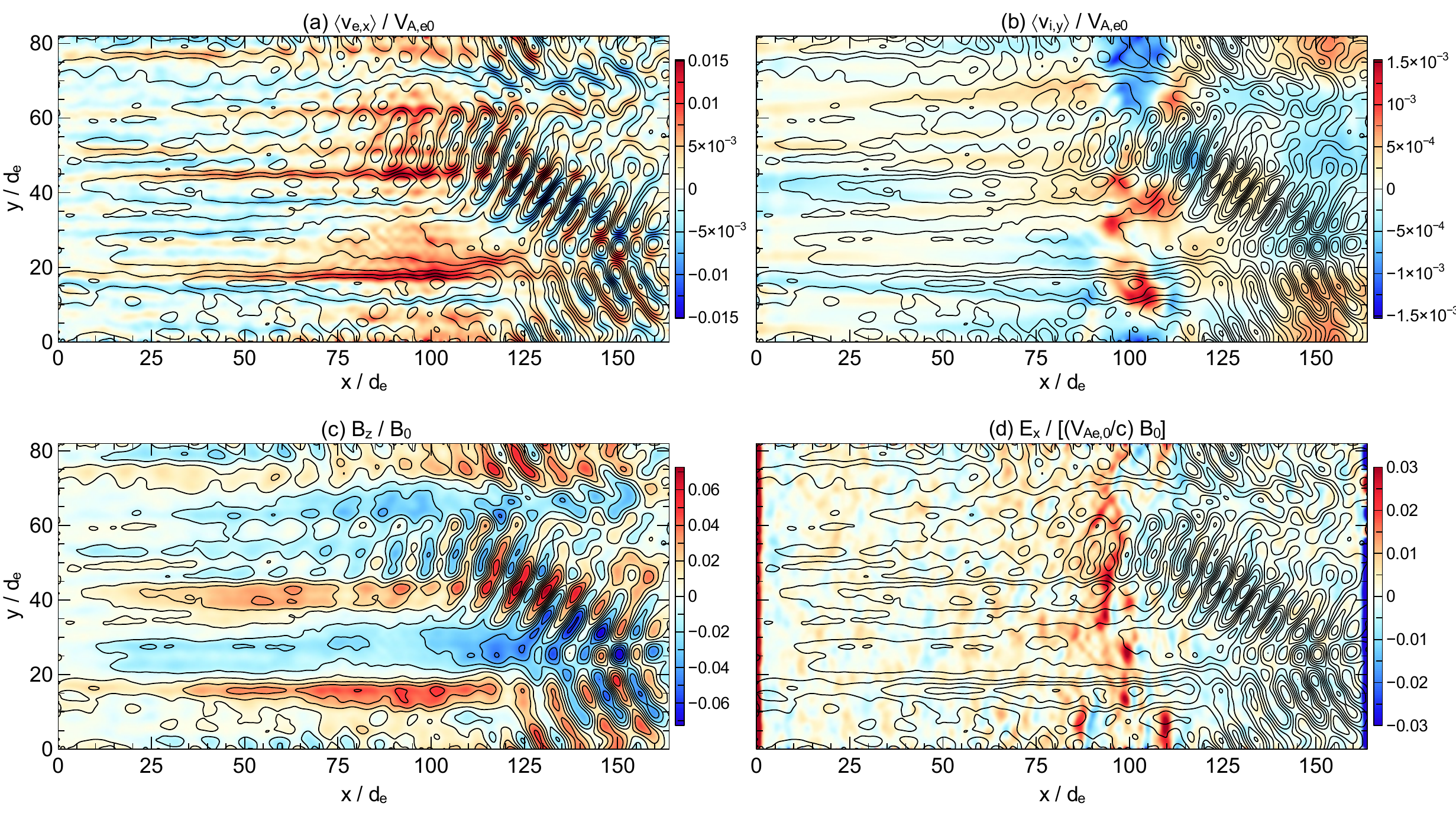}
\caption{2D plots showing the presence of whistlers and elongated magnetic structures at $t\Omega_{e0}=3640$ in the $\beta_{e0h}=1$ simulation. The schematic for current closure, producing the elongated magnetic structures seen for $x/d_{e} \lesssim 100$, is described in the text. Each plot has contours of out-of-plane $B_{z}$, representing in-plane current, overlaid. (a) Average parallel electron velocity $v_{ex}$. (b) Average perpendicular ion velocity $v_{iy}$. (c) $B_{z}$. (d) $E_{x}$.}
\label{fig:5}
\end{figure*}

\subsubsection{Weak heat flux instability}

In the $\beta_{e0h} \lesssim 1$ regime two factors lead to the stabilization of oblique whistlers: the phase speed of the whistlers and the impact of the DL on the electron distribution functions. Since the characteristic phase speed of the whistlers is the electron Alfv\'en speed $V_{A,e}$, in the low $\beta_{e0h}$ regime the phase speed exceeds the hot thermal speed. Since the hot/cold interface remains near $v_{x}=0$, only very long wavelength whistlers ($kd_{e}\ll 1$) could potentially resonate with electrons near the interface. Such long wavelength modes are not seen in our simulations. Thus, at low $\beta_{e0h}$ the whistler Landau resonance is at the wrong location in velocity space. The DL front also modifies the electron distribution in the vicinity of $v_{x}=0$ in both the hot and cold regions so as to enhance the stability of whistlers in this regime. Hot particles reflected by the DL move the hot/cold interface to the left in the hot region [fig. \ref{fig:3}(a)] so the Landau resonance is stabilizing. In figure \ref{fig:3}(b), on the other hand, the discontinuity in $f_{e}$ remains near $v_{x}=0$ but for whistlers with $kd_{e}\sim 1$ the Landau resonance again lies in the region $v_{x}>v_{Teh}$, well away from the location of the discontinuity.   

\subsection{Kinetic Alfv\'en Wave-like Structures} \label{sec:kaw}

\begin{figure*}
\plotone{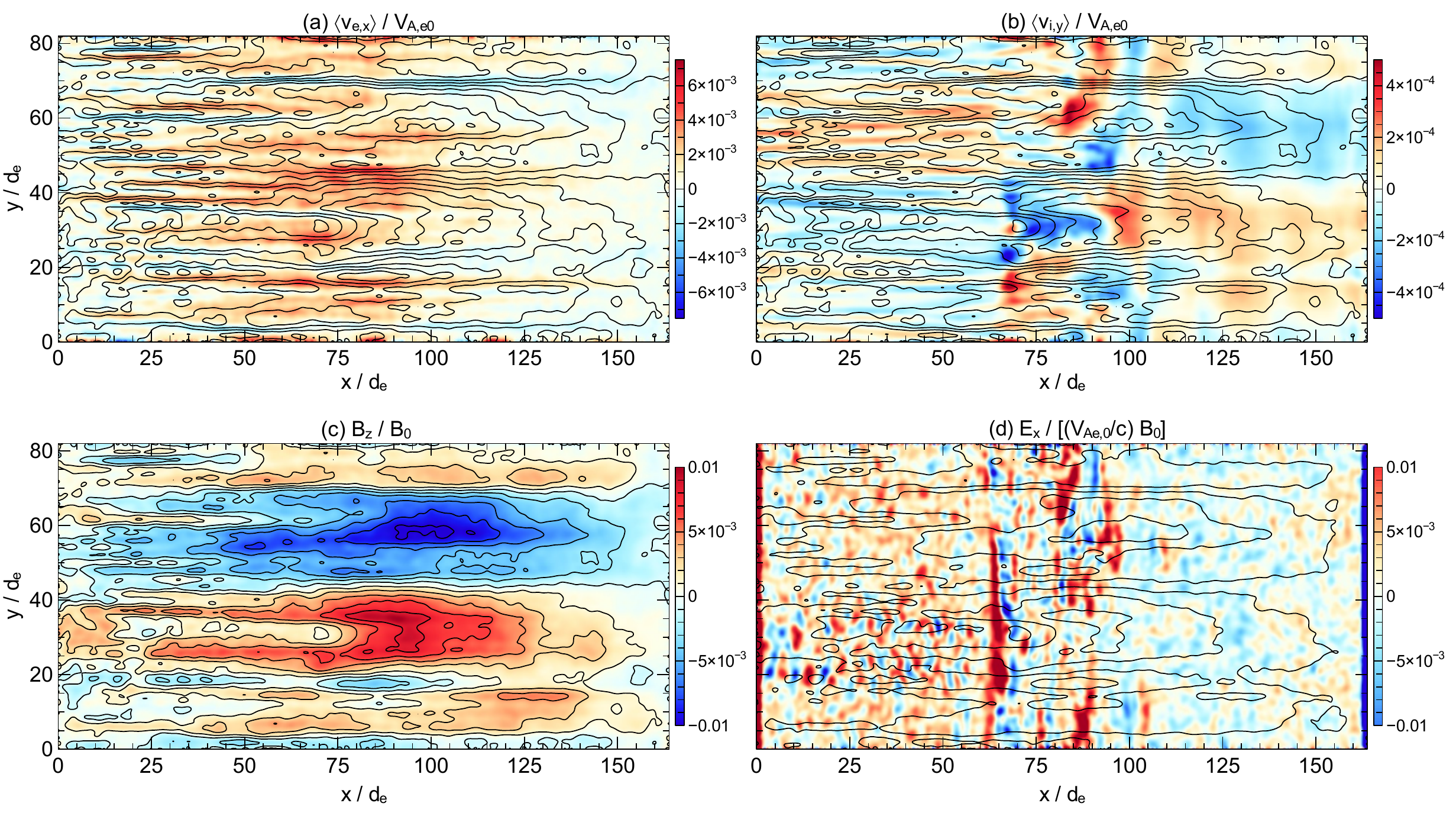}
\caption{Same as figure \ref{fig:5} but for the $\beta_{e0h}=1/4$ simulation at $t\Omega_{e0}=1820$.}
\label{fig:6}
\end{figure*}

The dynamics of the elongated magnetic structures $B_z$ in the $\beta_{e0h} \leq 1$ simulations can be described in terms of current loops formed in the $x,y$ plane. In figure \ref{fig:5} we present the components of these current loops in the $x,y$ plane ($v_{ex}$, $v_{iy}$ as well as $B_z$ and $E_x$). The region enclosed by these loops contains an out-of-plane $B_{z}$ [fig. \ref{fig:5}(c)]. Each plot in figure \ref{fig:5} has contours of $B_{z}$ overlaid, which correspond to the streamlines of the in-plane current, which can be written as $\mathbf{J}_{in-plane} \propto \hat{z} \times \nabla B_{z}$. We illustrate the nature of the structures by focusing on the oblate red structure in fig. \ref{fig:5}(c) roughly centered around $y/d_{e}=40$ in the hot region to the left of the DLs [seen in $E_{x}$, fig. \ref{fig:5}(d)]. The upper and lower sections of current that produce this particular structure result from electrons streaming along $\mathbf{B_{0}}=B_{0}\hat{x}$, seen in fig. \ref{fig:5}(a) as red ($y/d_{e} \sim 45$) and blue ($y/d_{e} \sim 36$) stripes in $\langle v_{ex} \rangle$ that closely follow the contours of in-plane current. Note that the electron current $J_{ex}=-en_{e}\langle v_{ex} \rangle$ differs in sign from $\langle v_{ex} \rangle$. The current is closed in the y direction by ions, which in fig. \ref{fig:5}(b) show a negative (blue) flow $\langle v_{iy} \rangle$ centered around $x/d_{e} \sim 18$ and a positive (red) flow near $x/d_{e} \sim 90$. In addition, oblique whistlers in the cold region to the right of the DLs may play some role in closing currents near the DLs. While the closure of in-plane currents by electrons and ions is typical of kinetic Alfv\'en waves (KAWs) \citep{Hasegawa1976,Hollweg1999}, $k_{\perp}\rho_{i} = \sqrt{m_{i}/m_{e}} k_{\perp}\rho_{e}$ is large so that the ions behave as unmagnetized particles in response to these disturbances (similar to ion heating by whistlers, section \ref{sec:whistler}). The waves therefore differ from the standard magnetized ion treatment of KAWs. 

The perpendicular wavelength $k_{y}$ of these structures is closely linked to the variation in the $y$ direction of the DL electrostatic potential $\Phi_{DL}$ (see $E_x$ in \ref{fig:5}(d)). We conjecture that it is the $y$ dependence (see \ref{sec:DL}) which causes return currents in the entire simulation domain to vary in $y$ and produces the elongated structures, but leave a more detailed exploration of the generation mechanism to future work.

\section{Discussion}

Using 2-D particle-in-cell simulations we have explored the physics of magnetized collisionless plasmas with a large, imposed electron heat flux for a range of $\beta_{e}$ from $1/4$ to $32$. The primary result is that a transition from heat flux suppression by whistler waves at high $\beta$ to that of electrostatic double layers (DLs) at low $\beta$ occurs [figure \ref{fig:4}(a)]. In this transition the whistler scaling for the parallel heat flux, $q_{e}/q_{0} \sim v_{Te} / \beta_{e}$, rolls over to a constant value $q_{e}/q_{0} \sim 0.3$ for the DL regime, where $q_{0}=n_{0}v_{Te}T_{e}$ is the free-streaming collisionless heat flux. This result could be compared with in-situ measurements of the solar wind heat flux at $1$ AU (e.g. \cite{Tong2018}).

The formation of the DLs comes about as a result of coupling between ions and electrons, which was not included in the earlier models of \cite{Roberg-Clark2018} and \cite{Komarov2018a}. The assumption of infinite mass ions in these models was nonetheless justifiable since DL formation is suppressed at high $\beta$ and ion damping has a small effect on the stability of whistlers. The DLs are generated by a return current linked to the strong imposed electron heat flux \citep{Li2012,Li2013,Li2014}. The DL system is inherently non-steady since the DL front propagates at finite speed against the direction of heat flux and periodically emits shocklets. Ion-electron coupling also includes ion heating downstream of the DLs at low $\beta$ and resonant heating of ions by whistlers at high $\beta_{e0h}$.

The heat-flux-driven whistlers saturate at large amplitude $\delta B \sim B_{0}$ in the high-$\beta$ regime and are suppressed at low $\beta$. The dominant magnetic perturbation at low $\beta$ was found to be an electron scale $k_{\perp}d_{e} \sim 1 \gg k_{\parallel}d_{e}$ mode with characteristics similar to a kinetic Alfv\'en wave. While the detailed generation mechanism for this mode has yet to be explored, we suspect it results from the inherent two-dimensionality of the DL conduction front and its associated electrostatic potential. This mode does not appear to impact electron thermal conduction but may be observable in $\beta \lesssim 1$ astrophysical plasmas such as the solar wind. Future measurements of heat fluxes and electromagnetic fields in the solar corona from the Parker Solar Probe mission could also be compared with our results.

A caveat of our model is that the simulation domain contains far fewer electron skin depths $d_{e}$ than than of a real astrophysical system. As a result, the imposed thermal gradient is much larger than that typically inferred in astrophysical environments such as the ICM \citep{Levinson1992}. On the other hand, our results on the limiting heat flux are surprisingly insensitive to the temperature gradient imposed in our simulations. Further, observations in the solar wind have confirmed the $1/\beta$ scaling of the limiting heat flux \citep{Tong2018} in spite of the fact that the gradient scale length of the temperature in the solar wind is very large compared with all kinetic scale lengths. The consequences of larger system sizes and weaker thermal gradients, e.g. the impact on saturation of and scattering by whistler fluctuations, are presently being explored. We also note that Coulomb collisions are not included in our model. The transition between the heat flux in systems with short collisional mean-free-paths and that of collisionless systems remains to be explored. Future work will address the impact of these results in fluid models of astrophysical plasmas.

\acknowledgments

The authors acknowledge support from NASA
grants NNX17AG27G, NNX14AF42G and NNN06AA01C as well as NSF Grant No. PHY1500460. This research used resources of the National Energy Research Scientific Computing Center, a DOE Office of Science User Facility supported by the Office of Science of the U.S. Department of Energy under Contract No.DE-AC02-05CH11231. Simulation data is available upon request.

 \software{p3d (Zeiler et al. 2002), IDL v. 8.2.2 (https://www.harrisgeospatial.com/docs/pdf/using.pdf), Veusz \: v. \: 3.0 \: (https://veusz.github.io/docs/manual.pdf)}

\appendix

\section{Whistler Free Energy}

Here we derive the available free energy for the heat flux instability at high $\beta$ when a hot electron half-Maxwellian distribution,

\begin{equation} \label{eqn:A1}
f_{h} = \frac{n_{0}}{\pi^{3/2}} \frac{ e^{-v^2/v_{Th}^2}}{v_{Th}^3}\theta(v_{x}),
\end{equation}
is isotropic in a frame moving with velocity $\mathbf{v_{p}}=v_{p}\mathbf{\hat{x}}$ of a large-amplitude whistler wave. $\theta(v_{x})$ is the Heaviside step function. For this calculation we have ignored the cold return current electron beam which contains half the total particles and enforces zero net current. We calculate the resulting distribution and free energy difference to lowest order in $v_{p}/v_{Th} \ll 1$. The isotropic distribution in the frame of the whistler is
\begin{equation} \label{eqn:A2}
f_{H}=\int_{0}^{\infty} dv_{x} \int_{0}^{\infty} dv_{\perp} 2\pi v_{\perp}   \delta \left(\sqrt{(v_{x}-v_{p})^2 + v^{2}_{\perp}}-v_{H} \right) f_{h}(v_{x},v_{\perp}),
\end{equation}
where $\delta(v)$ is the Dirac delta function and $v_{H}$ is the magnitude of the velocity measured from a coordinate system centered at $v_{x}=v_{p}$. Using 
\begin{equation}
\delta(g(x))=\sum_{i} \frac{\delta(x-x_{i})}{|g'(x_{i})|},
\end{equation}
where $x_{i}$ are the roots of the delta function and noting that the delta function has only one root if the $v_{\perp}$ integral is done first, we find, to lowest order in $v_{p}/v_{Th}$,

\begin{equation}
f_{H}=\frac{n_{0}}{2\pi^{3/2}v_{Th}^{3}}e^{-\frac{v^{2}_{H}}{v^{2}_{Th}}}\left( 1 + \frac{v_{p}}{v_{H}} - \frac{v_{p}v_{H}}{v^{2}_{Th}}  \right)
\end{equation}
for $v_{H} > v_{p}$. $f_{H}$ for $v_{H} < v_{p}$ gives 2nd order corrections to the energy and is ignored. The factor of $1/2$ comes from conservation of the number of particles between the initial and final states and the result of the integral (\ref{eqn:A2}) has been divided by $4\pi v^{2}_{H}$ to give a probability distribution. The energy of the distribution in the rest frame is
\begin{equation}
W_{f_{H}}=\frac{1}{2}m \int d^{3}v \: v^{2}f_{H}(v_{H}),
\end{equation}
where the rest frame velocity is $\mathbf{v} = \mathbf{v_{H}} + \mathbf{v_{p}}$. Computing the difference in energy between $f_{H}$ and $f_{h}$ (\ref{eqn:A1}) we find 
\begin{equation} \label{eqn:A6}
\Delta W = W_{f_{H}} - W_{f_{h}} = -\frac{n_{0}}{\sqrt{\pi}} \frac{v_{p}}{v_{Teh}}T_{eh},
\end{equation}
where higher-order terms in $v_{p}/v_{Th}$ were again neglected.

\bibliography{library}

\end{document}